\def\temp{1.34}%
\let\tempp=\relax
\expandafter\ifx\csname psboxversion\endcsname\relax
  \message{PSBOX(\temp) loading}%
\else
    \ifdim\temp cm>\psboxversion cm
      \message{PSBOX(\temp) loading}%
    \else
      \message{PSBOX(\psboxversion) is already loaded: I won't load
        PSBOX(\temp)!}%
      \let\temp=\psboxversion
      \let\tempp=\endinput
    \fi
\fi
\tempp
\let\psboxversion=\temp
\catcode`\@=11
%
%
\def\psfortextures{
\def\PSspeci@l##1##2{%
\special{illustration ##1\space scaled ##2}%
}}%
\def\psfordvitops{
\def\PSspeci@l##1##2{%
\special{dvitops: import ##1\space \the\drawingwd \the\drawinght}%
}}%
\def\psfordvips{
\def\PSspeci@l##1##2{%
\d@my=0.1bp \d@mx=\drawingwd \divide\d@mx by\d@my
\includegraphics{##1\space}}}%
\def\psforoztex{
\def\PSspeci@l##1##2{%
\special{##1 \space
      ##2 1000 div dup scale
      \number-\psllx\space \number-\pslly\space translate
}}}%
\def\psfordvitps{
\def\psdimt@n@sp##1{\d@mx=##1\relax\edef\psn@sp{\number\d@mx}}
\def\PSspeci@l##1##2{%
\special{dvitps: Include0 "psfig.psr"}
\psdimt@n@sp{\drawingwd}
\special{dvitps: Literal "\psn@sp\space"}
\psdimt@n@sp{\drawinght}
\special{dvitps: Literal "\psn@sp\space"}
\psdimt@n@sp{\psllx bp}
\special{dvitps: Literal "\psn@sp\space"}
\psdimt@n@sp{\pslly bp}
\special{dvitps: Literal "\psn@sp\space"}
\psdimt@n@sp{\psurx bp}
\special{dvitps: Literal "\psn@sp\space"}
\psdimt@n@sp{\psury bp}
\special{dvitps: Literal "\psn@sp\space startTexFig\space"}
\special{dvitps: Include1 "##1"}
\special{dvitps: Literal "endTexFig\space"}
}}%
\def\psfordvialw{
\def\PSspeci@l##1##2{
\special{language "PostScript",
position = "bottom left",
literal "  \psllx\space \pslly\space translate
  ##2 1000 div dup scale
  -\psllx\space -\pslly\space translate",
include "##1"}
}}%
\def\psforptips{
\def\PSspeci@l##1##2{{
\d@mx=\psurx bp
\advance \d@mx by -\psllx bp
\divide \d@mx by 1000\multiply\d@mx by \xscale
\incm{\d@mx}
\let\tmpx\dimincm
\d@my=\psury bp
\advance \d@my by -\pslly bp
\divide \d@my by 1000\multiply\d@my by \xscale
\incm{\d@my}
\let\tmpy\dimincm
\d@mx=-\psllx bp
\divide \d@mx by 1000\multiply\d@mx by \xscale
\d@my=-\pslly bp
\divide \d@my by 1000\multiply\d@my by \xscale
\at(\d@mx;\d@my){\special{ps:##1 x=\tmpx, y=\tmpy}}
}}}%
\def\psonlyboxes{
\def\PSspeci@l##1##2{%
\at(0cm;0cm){\boxit{\vbox to\drawinght
  {\vss\hbox to\drawingwd{\at(0cm;0cm){\hbox{({\tt##1})}}\hss}}}}
}}%
\def\psloc@lerr#1{%
\let\savedPSspeci@l=\PSspeci@l%
\def\PSspeci@l##1##2{%
\at(0cm;0cm){\boxit{\vbox to\drawinght
  {\vss\hbox to\drawingwd{\at(0cm;0cm){\hbox{({\tt##1}) #1}}\hss}}}}
\let\PSspeci@l=\savedPSspeci@l
}}%
%
%
\newread\pst@mpin
\newdimen\drawinght\newdimen\drawingwd
\newdimen\psxoffset\newdimen\psyoffset
\newbox\drawingBox
\newcount\xscale \newcount\yscale \newdimen\pscm\pscm=1cm
\newdimen\d@mx \newdimen\d@my
\newdimen\pswdincr \newdimen\pshtincr
\let\ps@nnotation=\relax
{\catcode`\|=0 |catcode`|\=12 |catcode`|
|catcode`#=12 |catcode`*=14
|xdef|backslashother{\}*
|xdef|percentother{
|xdef|tildeother{~}*
|xdef|sharpother{#}*
}%
\def\R@moveMeaningHeader#1:->{}%
\def\uncatcode#1{%
\edef#1{\expandafter\R@moveMeaningHeader\meaning#1}}%
\def\execute#1{#1}
\def\psm@keother#1{\catcode`#112\relax}
\def\executeinspecs#1{%
\execute{\begingroup\let\do\psm@keother\dospecials\catcode`\^^M=9#1\endgroup}}%
\def\@mpty{}%
\def\matchexpin#1#2{
  \fi%
  \edef\tmpb{{#2}}%
  \expandafter\makem@tchtmp\tmpb%
  \edef\tmpa{#1}\edef\tmpb{#2}%
  \expandafter\expandafter\expandafter\m@tchtmp\expandafter\tmpa\tmpb\endm@tch%
  \if\match%
}%
\def\matchin#1#2{%
  \fi%
  \makem@tchtmp{#2}%
  \m@tchtmp#1#2\endm@tch%
  \if\match%
}%
\def\makem@tchtmp#1{\def\m@tchtmp##1#1##2\endm@tch{%
  \def\tmpa{##1}\def\tmpb{##2}\let\m@tchtmp=\relax%
  \ifx\tmpb\@mpty\def\match{YN}%
  \else\def\match{YY}\fi%
}}%
\def\incm#1{{\psxoffset=1cm\d@my=#1
 \d@mx=\d@my
  \divide\d@mx by \psxoffset
  \xdef\dimincm{\number\d@mx.}
  \advance\d@my by -\number\d@mx cm
  \multiply\d@my by 100
 \d@mx=\d@my
  \divide\d@mx by \psxoffset
  \edef\dimincm{\dimincm\number\d@mx}
  \advance\d@my by -\number\d@mx cm
  \multiply\d@my by 100
 \d@mx=\d@my
  \divide\d@mx by \psxoffset
  \xdef\dimincm{\dimincm\number\d@mx}
}}%
%
\newif\ifNotB@undingBox
\newhelp\PShelp{Proceed: you'll have a 5cm square blank box instead of
your graphics (Jean Orloff).}%
\def\s@tsize#1 #2 #3 #4\@ndsize{
  \def\psllx{#1}\def\pslly{#2}%
  \def\psurx{#3}\def\psury{#4}
  \ifx\psurx\@mpty\NotB@undingBoxtrue
  \else
    \drawinght=#4bp\advance\drawinght by-#2bp
    \drawingwd=#3bp\advance\drawingwd by-#1bp
  \fi
  }%
\def\sc@nBBline#1:#2\@ndBBline{\edef\p@rameter{#1}\edef\v@lue{#2}}%
\def\g@bblefirstblank#1#2:{\ifx#1 \else#1\fi#2}%
{\catcode`\%=12
\xdef\B@undingBox{
\def\ReadPSize#1{
 \readfilename#1\relax
 \let\PSfilename=\lastreadfilename
 \openin\pst@mpin=#1\relax
 \ifeof\pst@mpin \errhelp=\PShelp
   \errmessage{I haven't found your postscript file (\PSfilename)}%
   \psloc@lerr{was not found}%
   \s@tsize 0 0 142 142\@ndsize
   \closein\pst@mpin
 \else
   \if\matchexpin{\GlobalInputList}{, \lastreadfilename}%
   \else\xdef\GlobalInputList{\GlobalInputList, \lastreadfilename}%
     \immediate\write\psbj@inaux{\lastreadfilename,}%
   \fi%
   \loop
     \executeinspecs{\catcode`\ =10\global\read\pst@mpin to\n@xtline}%
     \ifeof\pst@mpin
       \errhelp=\PShelp
       \errmessage{(\PSfilename) is not an Encapsulated PostScript File:
           I could not find any \B@undingBox: line.}%
       \edef\v@lue{0 0 142 142:}%
       \psloc@lerr{is not an EPSFile}%
       \NotB@undingBoxfalse
     \else
       \expandafter\sc@nBBline\n@xtline:\@ndBBline
       \ifx\p@rameter\B@undingBox\NotB@undingBoxfalse
         \edef\t@mp{%
           \expandafter\g@bblefirstblank\v@lue\space\space\space}%
         \expandafter\s@tsize\t@mp\@ndsize
       \else\NotB@undingBoxtrue
       \fi
     \fi
   \ifNotB@undingBox\repeat
   \closein\pst@mpin
 \fi
\message{#1}%
}%
%
%
\def\psboxto(#1;#2)#3{\vbox{%
   \ReadPSize{#3}%
   \advance\pswdincr by \drawingwd
   \advance\pshtincr by \drawinght
   \divide\pswdincr by 1000
   \divide\pshtincr by 1000
   \d@mx=#1
   \ifdim\d@mx=0pt\xscale=1000
         \else \xscale=\d@mx \divide \xscale by \pswdincr\fi
   \d@my=#2
   \ifdim\d@my=0pt\yscale=1000
         \else \yscale=\d@my \divide \yscale by \pshtincr\fi
   \ifnum\yscale=1000
         \else\ifnum\xscale=1000\xscale=\yscale
                    \else\ifnum\yscale<\xscale\xscale=\yscale\fi
              \fi
   \fi
   \divide\drawingwd by1000 \multiply\drawingwd by\xscale
   \divide\drawinght by1000 \multiply\drawinght by\xscale
   \divide\psxoffset by1000 \multiply\psxoffset by\xscale
   \divide\psyoffset by1000 \multiply\psyoffset by\xscale
   \global\divide\pscm by 1000
   \global\multiply\pscm by\xscale
   \multiply\pswdincr by\xscale \multiply\pshtincr by\xscale
   \ifdim\d@mx=0pt\d@mx=\pswdincr\fi
   \ifdim\d@my=0pt\d@my=\pshtincr\fi
   \message{scaled \the\xscale}%
 \hbox to\d@mx{\hss\vbox to\d@my{\vss
   \global\setbox\drawingBox=\hbox to 0pt{\kern\psxoffset\vbox to 0pt{%
      \kern-\psyoffset
      \PSspeci@l{\PSfilename}{\the\xscale}%
      \vss}\hss\ps@nnotation}%
   \global\wd\drawingBox=\the\pswdincr
   \global\ht\drawingBox=\the\pshtincr
   \global\drawingwd=\pswdincr
   \global\drawinght=\pshtincr
   \baselineskip=0pt
   \copy\drawingBox
 \vss}\hss}%
  \global\psxoffset=0pt
  \global\psyoffset=0pt
  \global\pswdincr=0pt
  \global\pshtincr=0pt 
  \global\pscm=1cm 
}}%
%
%
\def\psboxscaled#1#2{\vbox{%
  \ReadPSize{#2}%
  \xscale=#1
  \message{scaled \the\xscale}%
  \divide\pswdincr by 1000 \multiply\pswdincr by \xscale
  \divide\pshtincr by 1000 \multiply\pshtincr by \xscale
  \divide\psxoffset by1000 \multiply\psxoffset by\xscale
  \divide\psyoffset by1000 \multiply\psyoffset by\xscale
  \divide\drawingwd by1000 \multiply\drawingwd by\xscale
  \divide\drawinght by1000 \multiply\drawinght by\xscale
  \global\divide\pscm by 1000
  \global\multiply\pscm by\xscale
  \global\setbox\drawingBox=\hbox to 0pt{\kern\psxoffset\vbox to 0pt{%
     \kern-\psyoffset
     \PSspeci@l{\PSfilename}{\the\xscale}%
     \vss}\hss\ps@nnotation}%
  \advance\pswdincr by \drawingwd
  \advance\pshtincr by \drawinght
  \global\wd\drawingBox=\the\pswdincr
  \global\ht\drawingBox=\the\pshtincr
  \global\drawingwd=\pswdincr
  \global\drawinght=\pshtincr
  \baselineskip=0pt
  \copy\drawingBox
  \global\psxoffset=0pt
  \global\psyoffset=0pt
  \global\pswdincr=0pt
  \global\pshtincr=0pt 
  \global\pscm=1cm
}}%
%
\def\psbox#1{\psboxscaled{1000}{#1}}%
\newif\ifn@teof\n@teoftrue
\newif\ifc@ntrolline
\newif\ifmatch
\newread\j@insplitin
\newwrite\j@insplitout
\newwrite\psbj@inaux
\immediate\openout\psbj@inaux=psbjoin.aux
\immediate\write\psbj@inaux{\string\joinfiles}%
\immediate\write\psbj@inaux{\jobname,}%
%
%
\def\toother#1{\ifcat\relax#1\else\expandafter%
  \toother@ux\meaning#1\endtoother@ux\fi}%
\def\toother@ux#1 #2#3\endtoother@ux{\def\tmp{#3}%
  \ifx\tmp\@mpty\def\tmp{#2}\let\next=\relax%
  \else\def\next{\toother@ux#2#3\endtoother@ux}\fi%
\next}%
%
%
\let\readfilenamehook=\relax
\def\re@d{\expandafter\re@daux}
\def\re@daux{\futurelet\nextchar\stopre@dtest}%
\def\re@dnext{\xdef\lastreadfilename{\lastreadfilename\nextchar}%
  \afterassignment\re@d\let\nextchar}%
\def\stopre@d{\egroup\readfilenamehook}%
\def\stopre@dtest{%
  \ifcat\nextchar\relax\let\nextread\stopre@d
  \else
    \ifcat\nextchar\space\def\nextread{%
      \afterassignment\stopre@d\chardef\nextchar=`}%
    \else\let\nextread=\re@dnext
      \toother\nextchar
      \edef\nextchar{\tmp}%
    \fi
  \fi\nextread}%
\def\readfilename{\bgroup%
  \let\\=\backslashother \let\%=\percentother \let\~=\tildeother
  \let\#=\sharpother \xdef\lastreadfilename{}%
  \re@d}%
%
%
\xdef\GlobalInputList{\jobname}%
\def\psnewinput{%
  \def\readfilenamehook{
    \if\matchexpin{\GlobalInputList}{, \lastreadfilename}%
    \else\xdef\GlobalInputList{\GlobalInputList, \lastreadfilename}%
      \immediate\write\psbj@inaux{\lastreadfilename,}%
    \fi%
    \ps@ldinput\lastreadfilename\relax%
    \let\readfilenamehook=\relax%
  }\readfilename%
}%
\expandafter\ifx\csname @@input\endcsname\relax    
  \immediate\let\ps@ldinput=\input\def\input{\psnewinput}%
\else
  \immediate\let\ps@ldinput=\@@input
  \def\@@input{\psnewinput}%
\fi%
\def\nowarnopenout{%
 \def\warnopenout##1##2{%
   \readfilename##2\relax
   \message{\lastreadfilename}%
   \immediate\openout##1=\lastreadfilename\relax}}%
\def\warnopenout#1#2{%
 \readfilename#2\relax
 \def\t@mp{TrashMe,psbjoin.aux,psbjoint.tex,}\uncatcode\t@mp
 \if\matchexpin{\t@mp}{\lastreadfilename,}%
 \else
   \immediate\openin\pst@mpin=\lastreadfilename\relax
   \ifeof\pst@mpin
     \else
     \errhelp{If the content of this file is so precious to you, abort (ie
press x or e) and rename it before retrying.}%
     \errmessage{I'm just about to replace your file named \lastreadfilename}%
   \fi
   \immediate\closein\pst@mpin
 \fi
 \message{\lastreadfilename}%
 \immediate\openout#1=\lastreadfilename\relax}%
{\catcode`\%=12\catcode`\*=14
\gdef\splitfile#1{*
 \readfilename#1\relax
 \immediate\openin\j@insplitin=\lastreadfilename\relax
 \ifeof\j@insplitin
   \message{! I couldn't find and split \lastreadfilename!}*
 \else
   \immediate\openout\j@insplitout=TrashMe
   \message{< Splitting \lastreadfilename\space into}*
   \loop
     \ifeof\j@insplitin
       \immediate\closein\j@insplitin\n@teoffalse
     \else
       \n@teoftrue
       \executeinspecs{\global\read\j@insplitin to\spl@tinline\expandafter
         \ch@ckbeginnewfile\spl@tinline
       \ifc@ntrolline
       \else
         \toks0=\expandafter{\spl@tinline}*
         \immediate\write\j@insplitout{\the\toks0}*
       \fi
     \fi
   \ifn@teof\repeat
   \immediate\closeout\j@insplitout
 \fi\message{>}*
}*
\gdef\ch@ckbeginnewfile#1
 \def\t@mp{#1}*
 \ifx\@mpty\t@mp
   \def\t@mp{#3}*
   \ifx\@mpty\t@mp
     \global\c@ntrollinefalse
   \else
     \immediate\closeout\j@insplitout
     \warnopenout\j@insplitout{#2}*
     \global\c@ntrollinetrue
   \fi
 \else
   \global\c@ntrollinefalse
 \fi}*
\gdef\joinfiles#1\into#2{*
 \message{< Joining following files into}*
 \warnopenout\j@insplitout{#2}*
 \message{:}*
 {*
 \edef\w@##1{\immediate\write\j@insplitout{##1}}*
\w@{
\w@{
\w@{
\w@{
\w@{
\w@{
\w@{
\w@{
\w@{
\w@{
\w@{\string\input\space psbox.tex}*
\w@{\string\splitfile{\string\jobname}}*
\w@{\string\let\string\autojoin=\string\relax}*
}*
 \expandafter\tre@tfilelist#1, \endtre@t
 \immediate\closeout\j@insplitout
 \message{>}*
}*
\gdef\tre@tfilelist#1, #2\endtre@t{*
 \readfilename#1\relax
 \ifx\@mpty\lastreadfilename
 \else
   \immediate\openin\j@insplitin=\lastreadfilename\relax
   \ifeof\j@insplitin
     \errmessage{I couldn't find file \lastreadfilename}*
   \else
     \message{\lastreadfilename}*
     \immediate\write\j@insplitout{
     \executeinspecs{\global\read\j@insplitin to\oldj@ininline}*
     \loop
       \ifeof\j@insplitin\immediate\closein\j@insplitin\n@teoffalse
       \else\n@teoftrue
         \executeinspecs{\global\read\j@insplitin to\j@ininline}*
         \toks0=\expandafter{\oldj@ininline}*
         \let\oldj@ininline=\j@ininline
         \immediate\write\j@insplitout{\the\toks0}*
       \fi
     \ifn@teof
     \repeat
   \immediate\closein\j@insplitin
   \fi
   \tre@tfilelist#2, \endtre@t
 \fi}*
}%
\def\autojoin{%
 \immediate\write\psbj@inaux{\string\into{psbjoint.tex}}%
 \immediate\closeout\psbj@inaux
 \expandafter\joinfiles\GlobalInputList\into{psbjoint.tex}%
}%
%
%
%
\def\centinsert#1{\midinsert\line{\hss#1\hss}\endinsert}%
\def\psannotate#1#2{\vbox{%
  \def\ps@nnotation{#2\global\let\ps@nnotation=\relax}#1}}%
\def\pscaption#1#2{\vbox{%
   \setbox\drawingBox=#1
   \copy\drawingBox
   \vskip\baselineskip
   \vbox{\hsize=\wd\drawingBox\setbox0=\hbox{#2}%
     \ifdim\wd0>\hsize
       \noindent\unhbox0\tolerance=5000
    \else\centerline{\box0}%
    \fi
}}}%
%
\def\at(#1;#2)#3{\setbox0=\hbox{#3}\ht0=0pt\dp0=0pt
  \rlap{\kern#1\vbox to0pt{\kern-#2\box0\vss}}}%
%
\newdimen\gridht \newdimen\gridwd
\def\gridfill(#1;#2){%
  \setbox0=\hbox to 1\pscm
  {\vrule height1\pscm width.4pt\leaders\hrule\hfill}%
  \gridht=#1
  \divide\gridht by \ht0
  \multiply\gridht by \ht0
  \gridwd=#2
  \divide\gridwd by \wd0
  \multiply\gridwd by \wd0
  \advance \gridwd by \wd0
  \vbox to \gridht{\leaders\hbox to\gridwd{\leaders\box0\hfill}\vfill}}%
%
\def\fillinggrid{\at(0cm;0cm){\vbox{%
  \gridfill(\drawinght;\drawingwd)}}}%
%
%
\def\textleftof#1:{%
  \setbox1=#1
  \setbox0=\vbox\bgroup
    \advance\hsize by -\wd1 \advance\hsize by -2em}%
\def\textrightof#1:{%
  \setbox0=#1
  \setbox1=\vbox\bgroup
    \advance\hsize by -\wd0 \advance\hsize by -2em}%
\def\endtext{%
  \egroup
  \hbox to \hsize{\valign{\vfil##\vfil\cr%
\box0\cr%
\noalign{\hss}\box1\cr}}}%
%
\def\frameit#1#2#3{\hbox{\vrule width#1\vbox{%
  \hrule height#1\vskip#2\hbox{\hskip#2\vbox{#3}\hskip#2}%
        \vskip#2\hrule height#1}\vrule width#1}}%
\def\boxit#1{\frameit{0.4pt}{0pt}{#1}}%
\catcode`\@=12 
%
 \psfordvips   
\splitfile{\jobname}
\let\autojoin=\relax
\magnification=1200
\hsize=16.5truecm
\vsize=24truecm
\def\undertext#1{$\underline{\smash{\hbox{#1}}}$}
\hyphenation{Di-rac}
\hyphenation{fer-mi-ons}
\centerline{\bf Delicacies of the Mass Perurbation}
\centerline{\bf in the Schwinger Model on a Circle}
\vskip.5truecm
\centerline{M. B. Paranjape}
\centerline{ Laboratoire de physique nucl\'eaire, D\'epartement de physique,
Universit\'e de Montr\'eal,}
\centerline{C. P. 6128 succ. ``A", Montr\'eal,
Qu\'ebec, Canada, H3C 3J7}
\vskip.5truecm
\centerline{\bf Abstract}
\vskip.5truecm
\noindent
The Hilbert bundle for the massless fermions of the Schwinger model on a
circle,
over the space of gauge field configurations, is topologically non-trivial
(twisted).  The corresponding bundle for massive fermions is topologically
trivial (periodic).  Since the structure of the
fermionic Hilbert bundle changes discontinuously the possibility of perturbing
in the mass is thrown into doubt.  In this article, we show that a direct
application of the anti-adiabatic theorem of Low, allows the structure of the
massless theory to be dynamically preserved in the strong coupling limit,
${e\over m}>>1$.
This justifies the use of perturbation theory in the bosonized version of the
model, in this limit.
\vskip.5truecm
\centerline{\bf Introduction}
\vskip.5truecm\par\noindent
The massless Schwinger model on a circle was studied by Manton$^1$ and by
Hosotani and Hetrick$^2$.  It was pointed out that proper treatment of the
Wilson loop variable, the only dynamical degree of freedom in the gauge field,
was necessary to obtain the correct spectrum, that of a free massive scalar
theory.  The initial investigation of the massive Schwinger model$^3$, albeit
on
the line, often used perturbation in the mass.  In Ref. 4, the effects of
adding
a fermionic mass terms were investigated, primarily in the limit of large
fermion mass, or equivalently, weak coupling, ${e\over m}\rightarrow 0$.  We
direct the reader to Ref. 4 for a review and for more detailed references.  It
was found that the structure of the fermionic Hilbert bundle over the space of
gauge field configurations is drastically changed with the addition of a
fermionic mass.  This then indicates that the analyticity of the $m\rightarrow
0$ limit should be questioned.  In this article we examine the delicate nature
of the massless limit in more detail.  We show that it is indeed correct to
perturb in the mass about the twisted Hilbert bundle of the massless case.  The
dynamics induce transitions that cause the Hilbert bundle of the massive theory
to become twisted.  Basically, the fermionic energy levels that pass through an
exactly degenerate point in the massless case, pass through a quasi-degenerate
point in the massive case.  If the driving frequency is much larger than the
gap, the system can behave anti-adiabatically.  Of course if the frequency is
much smaller than the gap, it is the adiabatic theorem which governs the
dynamics, and the fermionic Hilbert bundle remains trivial.  Isler, Schmid and
Truegenburger$^5$ studied in detail the effects of mass on the Klein effect,
i.e., anomalous chiral charge production, due to {\it local} electric fields,
on
the line.  We are essentially considering anomalous chiral charge production
due
to {\it spatially constant} electric fields on a circle, which are not
permitted
on the line. \vskip.5truecm\par\noindent  The plan of this paper is as follows.
First we elaborate the structure of the Hilbert bundle for the massless and the
massive case.  Then we solve the relevant, external field Dirac equation to
demonstrate adiabatic behaviour of the weak coupling limit (large mass), and
the
anti-adiabatic behaviour of the strong coupling limit (small mass).  Finally we
show that the relevant driving frequencies indeed satisfy the appropriate
conditions in the two limits to permit perturbation about the corresponding
Hilbert bundles.   \vskip.5truecm
\centerline{\bf Hilbert Bundles}
\vskip.5truecm\par\noindent
We record here some results from Ref. 4, that we will find necessary.  The
classical Hamiltonian for the massive Schwinger model on a circle, of length
$2\pi L$, in the temporal gauge $A_0=0$, is given by
$$
{\cal H}=\int dx{1\over 2}(\dot v(x,t))^2+\Psi^\dagger
(x,t)\left(-i\gamma_5(\partial_x+iev(x,t))+ m\gamma^0\right)\Psi (x,t).\eqno(1)
$$
The only dynamical degree of freedom is the Wilson loop variable
$$
v={1\over 2\pi L}\int_0^{2\pi L}dxv(x).\eqno(2)
$$
Quantizing in the Schr\"odinger picture and eliminating the non-dynamical gauge
modes
with the Gauss law, yields the quantum Hamiltonian
$$
{\hat H}=-{1\over 4\pi L}{d^2\over dv^2}+{\hat
H}_F^0+evQ_5+Le^2v^2+{-1\over 2\pi L} \sum_{p\in Z\atop p\ne 0}{1\over
2}e^2L^2{j^0(p)j^0(-p)\over p^2}\eqno(3)
$$
where$^{4,6}$ ${\hat H}_F^0$ is the free Hamiltonian
$$
{\hat H}_F^0={1\over L}\sum_{p\in Z}\sqrt{p^2+(mL)^2}(a_p^\dagger
a_p+b_p^\dagger
b_p),\eqno(4)
$$
$Q_5$ is the axial charge and $j^0(p)$ is the charge density, see Ref. 4 or 6
for details.
We can diagonalize the fermionic Hamiltonian for the zero momentum sector
excluding the Coulomb energy term, exactly, which provides the vacuum energy.
Now comes an important divergence between the massive and the massless cases.
The massive case yields $$
{\hat H}_F={\hat H}_F^0+evQ_5+Le^2v^2=\sum_{p\in Z}\sqrt{({p\over
L}+ev)^2+m^2}\left(\bar a_p^\dagger  (v)\bar a_p(v)+\bar b_p^\dagger   (v)\bar
b_p(v)\right)-g(v) \eqno(5)
$$
with
$$
g(v)={-1\over  L}{2mL\over \pi}\sum_{n=1}^\infty{K_1(\pi nmL)\over n}(\cos
(2\pi
neLv)-1)\eqno(6)
$$
which is a smooth, periodic function of $v$, with period ${1\over eL}$.  The
limit
$m\rightarrow 0$ is not smooth.  We get
$$
{{\rm lim}\atop m\rightarrow 0}-g(v)={(eL)^2\over L}\left( v-{1\over
2eL}\right)^2-{1\over 4L} \eqno(7)
$$
on the domain $v\in [0,{1\over eL})$.  We could consider $-g(v)$ as periodic
with
discontinuous derivative at $0$, this is not, however, what we expect.  There
is no
reason to believe that the model is not smooth in $v$.  We must look in more
detail at
the massless case.  The limit $m\rightarrow 0$ of the Hamiltonian (5) is well
defined for each $v$ in the open set $(0,{1\over eL})$.  For $v=0$ or $v=
{1\over eL}$, however, the massless limit is not continuous.
\vskip.5truecm\par\noindent
The Dirac equation in the presence of a constant external gauge field is
trivially
solvable. For
$$
i\partial_t \Psi (x,t)=h(x,t)\Psi (x,t)=\left(-i\gamma_5(\partial_x
+iev)+m\gamma^0\right) \Psi (x,t)\eqno(8)
$$
the solutions are given by$^6$
$$
\eqalign{
\psi_\pm(x,p,v,m,L)&={e^{i{px\over L}}\over \sqrt{2\pi L}}\psi_\pm (p,v,m,L)\cr
&={e^{i{px\over L}}\over\sqrt{\left( 2(\sqrt{({p\over
L}-ev)^2+m^2})(\sqrt{({p\over L}-ev)^2+m^2}\mp ({p\over
L}-ev))\right)}}\times\cr
&\times\pmatrix{m\cr\cr\pm\sqrt {({p\over L}-ev)^2+m^2}-({p\over L}-ev)}}
\eqno(9)
$$
with corresponding energies $E_\pm (p,v,m,L)$,
$$
E_\pm (p,v,m,L)=\pm\sqrt {({p\over L}-ev)^2+m^2}.\eqno(10)
$$
\vskip.5truecm\par\noindent
The massless limit of the first quantized eigenfunctions for all $v\in
[0,{1\over eL})$
yields the correct, chiral eigenstates,
$$
{{\rm lim}\atop m\rightarrow
0}\psi^{\pm}(p,m,v,L)=\pmatrix{\theta(\pm(p+eLv))\cr
\theta(\mp(p+eLv))}=\cases {\pmatrix{1\cr 0}&chirality $+$\cr\pmatrix{0\cr
1}&chirality $-$},\eqno(11)
$$
while for $p+eLv=0$, for example for $v=0$, $p=0$, we have
$$
{{\rm lim}\atop m\rightarrow 0}\psi^{\pm}(p=0,m,v=0,L)={1\over\sqrt 2}
\pmatrix{1\cr\pm 1} \eqno(12)
$$
or for $v={1\over eL}$, $p=-1$, we have
$$
{{\rm lim}\atop m\rightarrow 0}\psi^{\pm}(p=-1,m,v={1\over eL},L)={1\over\sqrt
2}\pmatrix
{1\cr\pm 1}. \eqno(13)
$$
Due to the discontinuous behaviour in $v$, we do not expect the corresponding
annihilation and creation operators of the second quantized theory to be smooth
periodic
functions of $v$.  Indeed, as is well know by now, the massless Schwinger model
exhibits
the phenomena of spectral flow.  As $v$ increases from $0$ to ${1\over eL}$,
one positive
chirality state crosses zero energy from negative to positive, while one
negative
chirality state crosses zero energy from positive to negative.  The resulting
state
contains a chiral fermion--anti-fermion pair.  The limit from the massive
theory
automatically yields the state with this chiral pair removed, since the massive
theory exhibits no spectral flow.  Thus it is not surprising that periodicity
of
the massive theory is lost for the massless case.
\vskip.5truecm\par\noindent
In the massive case we have the unitary transformation
$$
\eqalign{
{\bar a}_p(v)&=V{\bar a}_{p+1}(v-{1\over eL})\cr
{\bar b}_p^\dagger (v)&=V^\dagger{\bar b}_{p+1}^\dagger (v-{1\over
eL})}\eqno(14)
$$
which relates annihilation and creation operators around the circle in $v$.
$V$
is actually independent of $v$ and relates the free operators in the following
way
$$
\eqalign{
a_p&=V^\dagger
\langle\psi^0_+(p)|\left(|\psi^0_+(p+1)\rangle a_{p+1}+|\psi^0_-(p+1)\rangle
b^\dagger_{p+1}\right)V\cr b^\dagger_p&=V^\dagger
\langle\psi^0_-(p)|\left(|\psi^0_+(p+1)\rangle a_{p+1}+|\psi^0_-(p+1)\rangle
b^\dagger_{p+1}\right)V,}\eqno(15)
$$
where $\psi^0_\pm (p)$ are the eigenfunctions at $v=0$.
In the massless limit this yields
$$
\eqalign{
a_p&=V^\dagger a_{p+1}\quad p\ne 0\cr
b^\dagger_p&=V^\dagger b^\dagger_{p+1}\quad p\ne 0}\eqno(16)
$$
but
$$
\eqalign{
a_0&=V^\dagger b^\dagger_1\cr
b^\dagger_0&=V^\dagger a_1.}\eqno(17)
$$
Thus for the massive case, $V$ does not mix annihilation operators with
creation
operators and vice versa, correspondingly, the vacuum is annihilated by the
same
annihilation operators at $v=0$ and at $v={1\over eL}$.  For the massless
case, however, it does mix annihilation and creation operators, exactly for
those
states which cross through zero, therefore the vacuum is not invariant.  Thus
in
the massive case, the vacuum is periodic, up to a phase, while in the massless
case it suffers a non-trivial unitary transformation which has the effect of
removing a chiral pair.  In either case the unitary transformation on the
fermionic Hilbert space is a representation of the topologically non-trivial
gauge transformation
$$
\psi(x,t)\rightarrow e^{i{x\over L}}\psi(x,t)\eqno(18)
$$
for the first quantized levels.  For the massless case we obtain a genuinely
twisted fermionic Hilbert bundle, while in the massive case the bundle is
topologically trivial, there are no topologically non-trivial complex line
bundles over the circle.
\vskip.5truecm\par\noindent
The inclusion of the Coulomb term is unimportant to the Hilbert bundle
structure
it does not involve any gauge field dependence.  The Coulomb term creates
fermion--anti-fermion pairs at zero total momentum which is, of course, of
great
importance for the full solution of the massless theory.
\vskip.5truecm\par\noindent
The discontinuity of the Hilbert bundles throws into
doubt the continuity of the massless limit of the massive Schwinger model, and
hence, perturbation theory in the mass.  Our aim here is to demonstrate that
mass perturbation is in fact without problem.  The kinematics of the massless
theory are reproduced by the dynamics of the massive theory, allowing for
smooth
behaviour as the mass tends to zero.  We will examine this in complete detail
in
the next section at the first quantized level.
\vskip.5truecm
\centerline{\bf First Quantized Theory}
\vskip.5truecm\par\noindent
The solutions (9) of the Dirac equation (8) have energies distributed along
the positive and negative branches of a hyperbola, at the integers, $p$,
shifted
by $eLv\in (0,1)$, see Figure 1.  As $v$ varies from $0$ to ${1\over eL}$,
the energies simply
permute by shifting along the hyperbola, the negative and positive branches
maintaining their identity.   In the massless limit, the hyperbola assumes its
asymptotes, which cross at $p=0$.  Chirality, the eigenvalue of $\gamma_5$, is
a
conserved quantum number in this case and the states now maintain their
identities along the lines $E={p\over L}$ for positive chirality and along
$E=-{p\over L}$ for  negative chirality.  Evidently negative energy
states and positive energy states do not maintain their own identities as $v$
increases from $0$ to ${1\over eL}$.
\centinsert{\pscaption{
\boxit{\psboxscaled{800}{delipics.ps}}}
{{\bf Figure 1}: Spectrum for the massive and massless cases, with $eLv=.2$,
$m=2.5$.}}
\vskip.5truecm\par\noindent
The topologically non-trivial gauge transformation (18) relates the wave
function at $v=0$ and $v={1\over eL}$.  This gauge transformation exactly
permutes the wave functions, for a given integer $p$ with that defined for the
integer $p+1$, along the branches of the hyperbola in the massive case,
while along the asymptotes in the massless case.  The corresponding unitary
transformation at the second quantized level (15) mixes creation and
annihilation
operators in the massless theory, but does not in the massive theory.  Hence
the
massive vacuum is invariant, while the massless vacuum is not.
\vskip.5truecm\par\noindent
Consider now the Dirac equation for a time
dependent, spatially constant external field $v(t)$.  We have
$$ i\partial_t
\Psi (x,t)=\left(-i\gamma_5(\partial_x +iev(t))+m\gamma^0\right) \Psi
(x,t)\eqno(19)
$$
Taking the spatial Fourier transform we see that the system separates into an
infinite set of independent two level systems indexed by the spatial momentum
$p$,
$$
i\partial_t \Psi (p,t)=\left(\gamma_5({p\over L}+ev(t) )+m\gamma^0\right) \Psi
(p,t)
\eqno(20)
$$
The time dependent instantaneous hamiltonian can be diagolnalized as follows
$$
\left(\gamma_5({p\over L}+ev(t) )+m\gamma^0\right)\psi_\pm
(p,v(t),m,L)=\pm\sqrt{({p\over L}+ev(t) )^2+m^2}
\psi_\pm (p,v(t),m,L)\eqno(21)
$$
with $\psi_\pm (p,v(t),m,L)$ given exactly as in (9).  Then expanding $\Psi
(p,t)$ in terms of the instantaneous eigenstates  $\psi_\pm (p,v(t),m,L)$
$$
\Psi (p,t)=a_+\psi_+ (p,v(t),m,L)+a_-\psi_- (p,v(t),m,L)\eqno(22)
$$
gives the two level system
$$
\left(i\partial_t-\left(\sqrt{({p\over L}+ev(t)
)^2+m^2}\sigma^3+{me\over2}{{\dot v(t)}\over ({p\over
L}-ev(t))^2+m^2}\sigma^2\right)\right)\pmatrix {a_+\cr a_-}=0,\eqno(23)
$$
where the $\sigma^i$ are the Pauli matrices.
Remarkably this system of equations can be exactly solved for certain time
dependence
$v(t)$, which we will use below.  The first term, proportional to $\sigma^3$,
is just the
instantaneous energy, while the second term proportional to $\sigma^2$ can
effect
transitions between instantaneous positive and negative energy states.
\vskip.5truecm\par\noindent
We can without loss of generality we consider the case $p=0$ as each two level
system behaves in the same way.  Now we imagine $v$ starting at large negative
values and increasing at a typical rate to large positive values.  Then for
$|ev|>>m$ we have approximately
$$
i\partial_t-{1\over L}\left( |evL|\sigma^3+\left({m\over ev}\right){L{\dot
v}(t)\over 2v}\sigma^2\right)\pmatrix{a_+\cr a_-}\approx  0.\eqno(24)
$$
Thus for ${L{\dot v}(t)\over v}\le o(1)$ the first term dominates and
transitions are negligible, we are in the adiabatic regime for either strong or
weak
coupling.  For $|ev|<<m$, however, we have approximately
$$
i\partial_t-{1\over L}\left( mL\sigma^3+{e\over m}{L{\dot v}(t)\over
2}\sigma^2\right)\pmatrix{a_+\cr a_-}\approx  0.\eqno(25)
$$
Now we obtain the adiabatic limit for ${L{\dot v}(t)\over v}<<{m\over ev}$.
Then in the strong coupling limit, ${m\over e}\rightarrow 0$, this inequality
can
be violated.  Once we obtain the situation ${m\over ev}<<{L{\dot v}(t)\over v}$
as ${m\over e}\rightarrow 0$, the adiabatic theorem does not apply.
Indeed it is the anti-adiabatic theorem of Low$^7$ which applies here, the
system behaves anti-adiabatically, the probability of transition instead of
being
negligible, becomes arbitrarily close to one.
\vskip.5truecm\par\noindent
We can elaborate the situation by taking the following time dependence for
$v(t)$,
$$
v(t)=\cases {-{m\over
e}{\alpha\over\sqrt{1-\alpha^2}}&$t<-{\alpha\over\omega}$\cr
             {m\over e}{\omega t\over\sqrt{1-(\omega
t)^2}}&$-{\alpha\over\omega}\le
              t\le{\alpha\over\omega}$\cr
              {m\over e}{\alpha\over\sqrt{1-\alpha^2}}&${\alpha\over\omega}<t$}
\eqno(26)
$$
with $\alpha$ some fixed dimensionless parameter between $0$ and $1$.  We get
$$
\eqalign{
&i\partial_t\pmatrix{a_+\cr a_-}=\cr
&-\left( {m\theta
(-t-{\alpha\over\omega})\over\sqrt{1-\alpha^2}}\sigma^3+{m\left(\theta
(t+{\alpha\over\omega})-\theta (t-{\alpha\over\omega})\right)\over
\sqrt{1-(\omega
t)^2}}\left(\sigma^3+{\omega\over 2m}\sigma^2\right)+{m\theta
(t-{\alpha\over\omega})\over\sqrt{1-\alpha^2}}\sigma^3\right)}\pmatrix{a_+\cr
a_-}   \eqno(27)
$$
which has the solution
$$
\pmatrix{a_+(t)\cr
a_-(t)}=\left(\cases{e^{-i(t+{\alpha\over\omega}){m\over\sqrt{1-\alpha^2}}\sigma^3}
&$t<-{\alpha\over\omega}$\cr
e^{-i{m\over\omega}\left(\sin^{-1}(\omega
t)+\sin^{-1}(\alpha)\right)\left(\sigma^3+{\omega\over 2m}\sigma^2\right)}&
$-{\alpha\over\omega}\le t\le{\alpha\over\omega}$\cr
e^{-i(t+{\alpha\over\omega}){m\over\sqrt{1-\alpha^2}}\sigma^3}
e^{-i{m\over\omega}\left(\sin^{-1}(\alpha)\right)\left(\sigma^3+{\omega\over
2m}\sigma^2\right)}&${\alpha\over\omega}<t$}\right)\pmatrix{a_+(-{\alpha\over\omega})\cr
a_-(-{\alpha\over\omega})},\eqno(28)
$$
with initial condition specified at $t=-{\alpha\over\omega}$.  For
$|t|>{\alpha\over\omega}$ there are no transitions, $v(t)$ is constant.  All
transitions occur between $-{\alpha\over\omega}$ and ${\alpha\over\omega}$,
thus
the transiton probability is determined by
$$
\pmatrix{a_+({\alpha\over\omega})\cr
a_-({\alpha\over\omega})}=e^{-i{2m\over\omega}\sin^{-1}(\alpha)\left(\sigma^3+{\omega\over
2m}\sigma^2\right)}\pmatrix{a_+(-{\alpha\over\omega})\cr
a_-(-{\alpha\over\omega})}.\eqno(29)
$$
As ${\omega\over m}\rightarrow 0$, we obtain the adiabatic limit,
$$
\pmatrix{a_+({\alpha\over\omega})\cr
a_-({\alpha\over\omega})}\rightarrow
e^{-i{2m\over\omega}\sin^{-1}(\alpha)\sigma^3}\pmatrix{a_+(-{\alpha\over\omega})\cr
a_-(-{\alpha\over\omega})}, \eqno(30)
$$
and there are negligible transitions.  As ${\omega\over m}\rightarrow\infty$,
however, we
get
$$
\pmatrix{a_+({\alpha\over\omega})\cr
a_-({\alpha\over\omega})}\rightarrow
e^{-i\sin^{-1}(\alpha)\sigma^2}\pmatrix{a_+(-{\alpha\over\omega})\cr
a_-(-{\alpha\over\omega})}. \eqno(31)
$$
Now if we further consider the massless limit (strong coupling), i.e., ${m\over
e}\rightarrow 0$, we must adjust $\alpha$ so that the change in
$ev$ is much larger than $m$, $e\Delta
v=e\left(v\left({\alpha\over\omega}\right)-
v\left(-{\alpha\over\omega}\right)\right)>>m$.  But
$$
e\Delta v=2{m\over
e}{\alpha\over\sqrt{1-\alpha^2}}>>m\eqno(32)
$$
implies we must take $\alpha\rightarrow 1$ as ${m\over e}\rightarrow 0$.
Hence  $\sin^{-1}(\alpha)\rightarrow{\pi\over 2}$ and
$$
{{\rm lim}\atop{{\omega\over m}\rightarrow\infty\atop{e\over
m}\rightarrow\infty}}
\pmatrix{a_+({\alpha\over\omega})\cr
a_-({\alpha\over\omega})}=e^{-i\sigma^2}\pmatrix{a_+(-{\alpha\over\omega})\cr
a_-(-{\alpha\over\omega})}=\pmatrix{-a_-(-{\alpha\over\omega})\cr
a_+(-{\alpha\over\omega})},\eqno(33)
$$
i.e., the transition probability is one.  The interpretation of this result is
quite straightforward.  The instantaneous fermionic eigenstates are essentially
chiral eigenstates at $t=-{\alpha\over\omega}$.  Then if $t$ increases to
${\alpha\over\omega}$, with $\omega\rightarrow\infty$, the sudden approximation
becomes relevant.  The eigenstates are essentially frozen as they were at
$t=-{\alpha\over\omega}$, but now continue to evolve according to the dynamics
of the Hamiltonian at $t={\alpha\over\omega}$.  Hence chirality is preserved.
A
previously negative energy, chirality plus wavefunction now evolves as a
positive energy, chirality plus wave function while a previously positive
energy, chirality minus wavefunction now evolves as a negative energy,
chirality
minus  wave function.  This is exactly as what occurs for the massless theory.
Therefore even though the Hilbert bundle of the massive fermions over the space
of gauge fields is  topologically trivial, it is appropriate to perturb about
the topologically twisted Hilbert bundle of the massless theory.
\vskip.5truecm
\centerline{\bf Driving Frequencies}
\vskip.5truecm\par\noindent
It remains to verify that the dynamics of the gauge fields do indeed satisfy
the appropriate condition on ${\omega\over m}$ in the strong and weak coupling
limit.  As elaborated in Ref. 4, the problem for the  single, dynamical gauge
degree of freedom in the weak coupling limit, ${e\over m}\rightarrow 0$,
corresponds to the one dimensional quantum mechanics system on a circle of
circumference $1\over eL$,
$$
({-1\over  4\pi L}{d^2\over  dv^2}-g(v))\psi (v)={\cal E}^0\psi (v),\eqno(34)
$$
with $g(v)$ given in (6).  Even this problem is not exactly solvable, however,
we
can extract enough relevant information in approximation.  Truncating the
series
at the first term, we obtain $$
-g(v)\approx -{4mL\over L\pi}K_1(2\pi mL)sin^2(\pi eLv)\eqno(35)
$$
which has is minimum at $v={1\over 2eL}$.  The harmonic approximation here
gives
$$
-g(v)\approx -{4mL\over L\pi}K_1(2\pi mL)\left(1-(\pi eL)^2(v-{1\over
2eL})^2+\cdots\right).\eqno(36)
$$
This harmonic approximation has for its ground state wave function
$$
\psi_0(v)\sim{\cal N}e^{-{2\pi L2e\sqrt{mlK_1(2\pi mL)}\over 2} (v-{1\over
2eL})^2} \eqno(37)
$$
which has a spread $\sigma$ approximately given by
$$
\eqalign{
\sigma &\sim{1\over (2\pi eL\sqrt{mlK_1(2\pi mL)})^{1\over 2}}\cr
&={2\over \pi e LS_0^{1\over 2}}}.\eqno(38)
$$
$S_0$ is the Euclidean action of the classical instanton which mediates
tunneling
around the circle, see Ref. 4 for details.  In the weak coupling limit,
$S_0\rightarrow\infty$.  Thus the ratio of $\sigma$ to the size of the circle,
$$
{\sigma\over\left({1\over eL}\right)}={2\over \pi S_0^{1\over 2}}\rightarrow 0.
\eqno(39)
$$
Therefore the spread of the wave function is negligible compared to the size of
the circle and the harmonic approximation is reasonable.  The relevant
frequency
then is just
$$
\omega = 2e\sqrt{mlK_1(2\pi mL)},\eqno(40)
$$
which gives a reasonable measure of $\dot v$.  Hence
$$
{{\dot v}\over m}\sim 2{e\over m}\sqrt{mlK_1(2\pi mL)}<<1\eqno(41)
$$
for weak coupling and the adiabatic treatment is valid.  The relevant Hilbert
bundle structure is the periodic, topologically trivial structure associated to
the weak coupling limit.
\vskip.5truecm\par\noindent
In the strong coupling limit, i.e., near the massless theory, the harmonic
approximation about the minimum of the potential fails to be reasonable.  The
spread of the ground state wave function surpasses the size of the circle
$$
{\sigma\over\left({1\over eL}\right)}\rightarrow\infty.\eqno(42)
$$
The size of the circle is now the determining constraint to find the typical
value of  $\dot v$.  A free particle on a circle of circumference ${1\over
eL}$, with kinetic term the same as that in (34),  has energy levels
$$
{\cal
E}_n={(2\pi eL)^2\over 4\pi L}n^2.\eqno(43)
$$
The corresponding frequency is
$$
\omega^\prime = {(2\pi eL)^2\over 4\pi L}.\eqno(44)
$$
Thus
$$L{\dot v}={(2\pi eL)^2\over 4\pi}>>{m\over e},\eqno(45)
$$
since  ${m\over e}\rightarrow 0$ in the strong coupling limit.  Then it is
clear
that the anti-adiabatic treatment is relevant and it is incorrect to neglect
the
dynamical twisting of the fermionic Hilbert bundle.  Since the transition
probability is one in the anti-adiabatic limit, we recover the Hilbert bundle
structure of the massless theory.  As shown in Ref. 1, the bosonization process
inherently takes into account the twisted nature of the fermionic Hilbert
bundle.  Hence analyses which involve perturbation theory in the mass are
properly done in the bosonized version, such as in Ref. 3.
\vskip.5truecm\par\noindent
In conclusion, we have shown that it is correct to perturb in the mass, about
the
topologically non-trivial (twisted) fermionic Hilbert bundle structure of the
massless theory, even though the Hilbert bundle structure of the massive
Schwinger model on a circle is topologically trivial (periodic).  The twisted
structure is dynamically maintained for small mass (strong coupling) following
the anti-adiabatic theorem of Low$^7$.
\vskip.5truecm
\centerline{\bf Acknowlegements}
\vskip.5truecm\par\noindent
We thank R. B. MacKenzie for
useful discussions and the McGill High Energy Physics Theory Group, the
preparation of a seminar for which, illicited the preceding enquiry.  This work
supported in part by NSERC of Canada and FCAR of Qu\'ebec.
\vskip.5truecm
\centerline{\bf References}
\vskip.5truecm
\par\noindent
1.  N. S. Manton, Annals of Physics,{\bf 159}, 220, (1985).
\par\noindent
2.  Y. Hosotani and J. Hetrick, Phys. Lett. {\bf B230}, 88, (1989); Phys. Rev.
{\bf D38}, 2621, (1988).
\par\noindent
3.  S. Coleman, R. Jackiw, L. Susskind, Annal of Physics, {\bf 93}, 267,
(1975);S. Coleman, ibid, {\bf 101}, 239, (1976).
4.  M. B. Paranjape and R. Ross, Phys. Rev. {\bf D}, in press.
\par\noindent
\par\noindent
5.  K. Isler, C. A. Truegenberger and C. Schmid, Nucl. Phys.,{\bf B314}, 269,
(1989).
\par\noindent
6.  M. B. Paranjape, Phys. Rev. {\bf D40}, 540, (1989).
\par\noindent
7.  F. Low, Phys. Rev. Lett. {\bf 63}, 2322, (1989).
\autojoin
\end